\documentclass[aps,prl,groupedaddress,]{revtex4}
\usepackage{latexsym}
\usepackage[dvips]{graphicx}

\newcommand{\eq}{\begin{equation}}
\newcommand{\feq}{\end{equation}}
\newcommand{\eqn}{\begin{eqnarray}}
\newcommand{\feqn}{\end{eqnarray}}
\newcommand{\arr}{\begin{eqnarray*}}
\newcommand{\farr}{\end{eqnarray*}}
\newcommand{\beq}{\begin{equation}}
\newcommand{\eeq}{\end{equation}}
\newcommand{\bea}{\begin{eqnarray}}
\newcommand{\eea}{\end{eqnarray}}

\def\beq{\begin{equation}}
\def\eeq{\end{equation}}
\def\feq{\end{equation}}
\def\bea{\begin{eqnarray}}
\def\eea{\end{eqnarray}}
\def\bc{\begin{displaymath}}
\def\ec{\end{displaymath}}
\def\lb{\label}

\def\be{\beta}
\def\g{\gamma}

\def\de{\delta}

\def\la{\lambda}

\def\Om{\Omega}

\def\lb{\label}

\begin{document}


\title{  Two-dimensional description of  $D$-dimensional static black holes with
pointlike sources }

\author{Mariano Cadoni}
\email{mariano.cadoni@ca.infn.it}

\affiliation{Dipartimento di Fisica,
Universit\`a di Cagliari, and INFN, Sezione di Cagliari, Cittadella
Universitaria 09042 Monserrato, Italy}
\author{Salvatore Mignemi}
\email{ smignemi@vaxca1.unica.it}
\affiliation{Dipartimento di  Matematica, Universit\`a di Cagliari,
viale Merello 92, 09123 Cagliari, Italy, and INFN, Sezione di Cagliari}


\begin{abstract}
We derive two-dimensional (2D) solutions of a generic dilaton gravity model
coupled with matter, which describe $D$-dimensional static black holes
with pointlike sources.  The equality between the mass $M$ of the
$D$-dimensional gravitational solution and the mass $m$ of the source
can be preserved also at the level of the 2D gravity model.

\end{abstract}


\maketitle

Two-dimensional (2D) models of gravity have been used in various contexts
as  an effective description  of  $D$-dimensional gravity.
In the case of pure gravity
the 2D dilaton gravity  model (see e.g Ref. \cite{Grumiller:2002nm} and
references therein)
can be integrated exactly and it can  be used
to give an effective description of the radial modes  of $D$-dimensional
gravity at the classical, semiclassical and full quantum level.
It is almost impossible to give a comprehensive  list of all the problems of
black hole  physics that  have been tackled  using 2D  dilaton gravity models.
The Hawking radiation and the information puzzle \cite{Callan:1992rs} for black
holes, the
microscopic interpretation of the black hole entropy \cite{Cadoni:1999ja} and the
construction of acoustic analogues of gravitational black holes \cite{Cadoni:2004my} 
are only a few  of many examples.

Unfortunately, many nice features of the 2D model are spoiled
by  the presence of matter. The existence and uniqueness of the
solutions  are consequence of a  2D Birkhoff's theorem
\cite{Louis-Martinez:1993cc,Cavaglia:1998uc},  which
has been shown to hold for pure 2D dilaton gravity but in general does
not hold when matter is present.
This is a strong limitation of   2D models. It is therefore
important to look for general solutions of matter-coupled dilaton
gravity models.

In this letter we will consider a generic 2D dilaton gravity model coupled with
a pointlike source. The model gives an effective description of
$D$-dimensional static black holes  with pointlike sources.
Moreover, gravitational models in which the matter contribution is
represented by $\delta$-like terms emerge in brane world scenarios
\cite{Randall:1999vf,Brax:2004xh}. 
We will derive the general solution  of the model  and show that
choosing appropriately the integration constants, the equality between
gravitational and inertial mass  can be preserved at the 2D level.

Let us consider  generic $D$-dimensional Einstein gravity coupled  to
matter fields and a  pointlike source of mass $m$.  The action
is \cite{Wei}
\beq\lb{ea}
A_D=\frac{1}{16\pi G_{D}}\int d^Dx\sqrt{-g}\ R- \int
d^Dx\sqrt{-g}\ {\cal L}_{MF}- m\int
dt\sqrt{-g_{ij}\frac{dx^{i}}{dt}\frac{dx^{j}}{dt}},
\eeq
where  $G_{D}$ is the $D$-dimensional  Newton constant and ${\cal L}_{MF}$ is 
the Lagrangian for the matter fields (gauge fields, scalar fields etc).

It is well known that in absence of the pointlike source
$D$-dimensional spherically symmetric gravity allows for an effective
two-dimensional description in terms of  a 2D dilaton gravity
model. The 2D model is obtained by retaining only the radial modes in
the  action (\ref{ea}). After a suitable Weyl rescaling of the metric
one gets  the 2D action $A_{2}=\int d^{2}x\sqrt{-g}\left(\phi R
+\lambda^{2}V(\phi)\right).$
The scalar field $\phi$ (the dilaton) parametrizes the volume of the transverse
$(D-2)$-dimensional sphere, $\la$ is a parameter with dimensions of
length$^{-1}$ and the dilaton  potential  $V(\phi)$ depends
both on the dimension of the spacetime $D$ and on the form of the
matter field action ${\cal L}_{MF}$ (We shall only consider the case
in which the matter fields can be treated, from the 2D point of view,
as constant background solutions).

When the pointlike source is present the situation becomes more
involved and it is not a priori evident that a 2D description of the
gravity model (\ref{ea}) exists.  Two main problems arise.
First, the  $D$-dimensional stress-energy tensor for the  pointlike source
is proportional to $\delta^{D}(x-x(\tau))$. Solving the
$D$-dimensional field
equations in the case  of spherical symmetry is essentially equivalent
to find the  Green's functions of the Laplace operator  in $D-1$ dimensions.
Thus, to give a 2D description of the model (\ref{ea}) a
correspondence  between $(D-1)$-dimensional and one-dimensional
Green's functions
is required.
Assuming that this correspondence has been found, one is faced with a
second problem. Even though the mass $M$ of the $D$-dimensional solution
equals the mass $m$ of the the source, one may end up  with
a 2D  solution for which $M\neq m$.

Progress in solving the first problem has been achieved in Ref.\ \cite{Melis:2004bk}
(see also Refs. \cite{Mann:1989gh,Mann:1991ny,Leite:1996ff,Boehmer:2005it}).   
In that paper the one-dimensional
Green's functions
have been given as functions of $|r|$, $r$ being the spacelike
coordinate of the 2D spacetime, but
unfortunately,  the  solutions of the 2D gravity model are
characterized by $m\neq M$.

Retaining only the radial modes in the action (\ref{ea}) and after a
Weyl rescaling of the 2D metric needed to get rid of the kinetic terms
for the dilaton, one obtains a 2D model characterized by a
dilaton potential $V(\phi)$  and a coupling function $W(\phi)$,
\beq\lb{dg}
A=A_{G}+A_{M}= \frac{1}{2} \int d^{2}x\sqrt{-g}\left(\phi R
+\lambda^{2}V(\phi)\right)- m\ \int dt\,W(\phi)\sqrt{-g_{\mu\nu}\frac{dx^{\mu}}{dt}
\frac{dx^{\nu}}{dt}}\,.
\feq
The form of the coupling function $W(\phi)$ depends both on the
dimensionality $D$ of the spacetime and on the Weyl rescaling
performed on the
2D metric.

The field equations for the 2D dilaton gravity  model (\ref{dg}) are 
(we do not consider the motion of the source)
\bea\lb{fe}
&&R= -\la^{2} \frac{dV}{d\phi}+\frac{2m}{\sqrt{-g}}{dW\over
d\phi}\,\int d\tau \,\de^{2}(x-x(\tau)),\nonumber\\
&&-(\nabla_{\mu}\nabla_{\nu}\phi - g_{\mu\nu}\nabla^{2}\phi)-\frac{1}{2}
g_{\mu\nu}\la^{2}V=T_{\mu\nu}^{(M)},
\eea
where $\tau$ is the proper time  and
$T_{\mu\nu}^{(M)}$ is the stress energy tensor for the pointlike matter
source,
\beq\lb{et}
T_{\mu\nu}^{(M)}= \frac{m}{\sqrt{-g}}\,W(\phi)
\int d\tau\, \de^{2}(x-x(\tau)) u_{\mu} u_{\nu},
\feq
with $u^\mu$ the 2-velocity of the particle.

If we consider the field equations in vacuum $T_{\mu\nu}^{(M)}=0$,
we can define a scalar function
\cite{Mann:1992yv}
\beq\lb{mass}
M= \frac{1}{2\la}\left(\la^{2}\int V(\phi) d\phi -
(\nabla\phi)^{2}\right),
\feq
which is constant  by virtue of the field equations  and gives
the mass of the gravitational solution.

Adopting  static coordinates and  the Schwarzschild gauge,
\beq\lb{gf}
ds^{2}= -
    U(r) dt^{2}
    + U^{-1}(r) dr^{2},
\feq
and considering a source particle at rest at the origin, the field equations
(\ref{fe}) become,
\bea\lb{fem}
&&\frac{d^{2}U}{dr^{2}}=\lambda^{2}\frac{dV}{d\phi}
-2m{dW\over d\phi}\,\de(r),\nonumber\\
&&\frac{dU}{dr}\frac{d\phi}{dr}=\lambda^{2}V,\\
&&2U\frac{d^{2}\phi}{dr^{2}}+\frac{dU}{dr}\frac{d\phi}{dr}=\lambda^{2}V
-2mW\,\de(r)\nonumber,
\eea
The general solution of  Eqs.\ (\ref{fem})  is a generalization of that
found in Ref.\ \cite{Melis:2004bk}.  In  two spacetime dimensions the coordinate $r$ has
not the meaning of
a radial coordinate, so that one can take  $-\infty< r< \infty$.
The delta function singularity at $r=0$ can be generated writing the
solution as function of $|r|$,
\beq\lb{gs}
U=\frac{J(\phi)}{\sigma^{2}}-\gamma,\quad \phi=\sigma \la |r|+\beta,
\feq
where $J(\phi)=\int V(\phi)d\phi$ and $\sigma,\beta, \gamma$ are
integration constants.
Inserting Eqs.\ (\ref{gs}) into Eqs.\ (\ref{fem}), one easily finds that
(\ref{gs}) is a solution of the system (\ref{fem}) if  the integration
constants satisfy
\beq\lb{ic}
J(\beta)-\sigma^{2}\gamma= - \frac{m\sigma}{2\la} W(\beta),\quad
V(\beta)=-\frac{m\sigma}{\la} \frac {dW}{d\phi}(\beta).
\feq
In general the mass $M$ of the solution, calculated using Eq.\ (\ref{mass})
away from the source in the $r>0$ region, is not equal to the mass $m$
of the source. However, in 2D an additional freedom is present
compared to higher dimensions. In fact, the solution
(\ref{gs}) is parametrized by three integration constants constrained
by only two equations (\ref{ic}). The remaining freedom can be fixed
requiring $m=M$. The origin of this additional freedom can be traced
back to the peculiarity of 2D gravity: the radial coordinate  has no
natural normalization in two dimensions. More precisely, the field
equations (\ref{fem}) in absence sources are invariant under the rescaling
$r\to \Om r$, $U\to \Om^2 U$, with constant $\Om$. This is analogous to
the possibility of freely rescaling the time coordinate in generic
dimension. The invariance  implies the presence in the solution (\ref{gs})
of an  integration constant, $\sigma$, with no higher-dimensional analogue,
that can be fixed arbitrarily.  When sources are added, one must of course
rescale also $m$ to preserve the invariance.

Requiring $m=M$ and using equation (\ref{gs}) into Eq.\ (\ref{mass})
one finds
\beq\lb{ic1}
\gamma= \frac{2m}{\la\sigma^{2}}.
\feq

As an example, we consider the  case $V=a\phi^h$, $W=b\phi^k$ ($a,b,h,k$
are arbitrary real constants).
Using equations (\ref{gs}), (\ref{ic}) and (\ref{ic1}) one finds  the
solution
\beq\lb{f17}
U=\frac{a}{\sigma^{2}(h+1)}(\sigma\la|r|+\be)^{h+1}-\g,\qquad
\phi=\sigma\la|r|+\be,
\eeq
with
\bea\lb{f1}
\beta&=&B^{\frac{1}{(h+1)}} \left(\frac{m}{\la a}\right)^{\frac{1}{h+1}}\nonumber\\
\sigma&=&- \frac{1}{bk}B^{\frac{h-k+1}{h+1}} \left(\frac{m}{\la
a}\right)^{-\frac{k}{h+1}}\\
\gamma&=&2a b^{2} k^{2}B^{\frac{2(k-h-1)}{h+1}} \left(\frac{m}{\la
a}\right)^{\frac{2k+h+1}{h+1}}.\nonumber
\eea
where $B=(4k(h+1)/(2k-h-1)$. The solution exists for  every $B>0$.

A particularly interesting special case of the power-law potentials 
discussed above is the dimensional reduction of the four-dimensional
(4D) Schwarzschild solution.
Consider  the 4D Einstein action coupled to a point particle
of mass $m$ given by Eq.\ (\ref{ea}) with ${\cal L}_{MF}=0$,
and substitute the ansatz
\beq\lb{f15}
ds^{2}_{(4)}=ds^{2}_{(2)}+ \frac{2}{\lambda^{2}}\phi\, d\Omega^{2}_{2}.
\feq
After a Weyl rescaling of the 2D metric,
\beq\lb{wr}
g\to \frac{1}{\sqrt{2\phi}}\ g,
\eeq
one finds  a 2D  model of the form (\ref{dg})  with
\beq\label{f4}
V(\phi)= \left( 2\phi\right)^{-1/2}, \quad W(\phi)=\left(
2\phi\right)^{-1/4}.
\eeq
Using Eqs.\ (\ref{f17}) and (\ref{f1}), with
$h=-\frac{1}{2},k=-\frac{1}{4}, a=\frac{1}{\sqrt 2}, b= (2)^{-1/4}$,
one easily finds for $r>0$
\beq\lb{f16}
U= \sigma^{-2}\sqrt{2\phi}-\gamma,\quad \phi= \sigma \la
r+\beta,\quad \sigma= 2\sqrt{\frac{m}{\la}},\quad
\beta=\frac{m^{2}}{2\la^{2}},\quad \gamma=\frac{1}{2}.
\eeq

It is also straightforward to check that after going back to the original
Weyl frame, $g_{s}= g/\sqrt{2\phi}$, changing coordinates
\beq
t'={t\over \sigma}, \quad R^2=\frac{2}{\lambda^{2}}\phi= \frac{2}{\lambda^{2}}
\left(\sigma \lambda r +\beta\right),
\eeq
and setting $G= \lambda^{-2}$, the  2D solution (\ref{f16})
reduces to the 4D Schwarzschild solution, $ds^{2}= -(1- 2Gm/R)dt'^{2}
+(1-2Gm/R)^{-1}dR^{2}+ R^{2}d\Omega^{2}_{2}$.

In the same way one may discuss the dimensional reduction of more general 
solutions like Reissner-Nordstr\"om or Schwarzschild-anti de Sitter.


\end{document}